# Modeling Supply Chain Interaction and Disruption: Insights from Real-world Data and Complex Adaptive System

Jiawei Feng[1], Mengsi Cai[1], Fangze Dai[2], Tianci Bu[1], Xiaoyu Zhang[1], Huijun Zheng[1], Xin Lu[1,3*]

(1. College of Systems Engineering, National University of Defense Technology, Changsha, China;

2. School of Management, Zhejiang University, Hangzhou, China;

3. Flowminder, Karolinska Institute, Department of Global Public Health, Stockholm, Sweden)

**Abstract:** In the rapidly evolving automotive industry, Systems-on-Chips (SoCs) are playing an increasingly crucial role in enhancing vehicle intelligence, connectivity, and safety features. For enterprises whose business encompasses automotive SoCs, the sustained and stable provision and receipt of SoC relevant goods or services are essential. Considering the imperative for a resilient and adaptable supply network, enterprises are concentrating their efforts on formulating strategies to address risks stemming from supply chain disruptions caused by technological obsolescence, natural disasters, and geopolitical tensions. This study presents an open supply knowledge extraction and complement approach and build a supply chain network of automotive SoC enterprises in China, which incorporates cross-domain named entity recognition under limited information, fuzzy matching of firm entities, and supply relation inferring based on knowledge graph. Subsequently, we exhibit the degree and registered capital distribution across firms, and analyze the correlations between centrality metrics in the supply chain network. Finally, two interaction disruption models (IDMs), based on recovery capacity and risk transfer, are developed to elucidate the adaptive behaviors and effect of network disruptions under various business and attack strategies. This research not only aids in exploring the complexities of Chinese automotive SoC supply chain but also enriches our understanding of the dynamics of firm behavior in this crucial industry sector.

**Keywords**: supply chain, complex networks, interaction disruption model, cascade failure

## 1. Introduction

As global politics, economics, and culture undergo continuous transformation, the international trade environment has become increasingly sever along with uncertain factors growing. Supply chain disruption risks constantly threaten the stability and efficiency of global semiconductor trade markets. Consequently, enhancing supply chain's ability to predict and withstand risks posed by natural disasters, social security issues, public health crises, and political conflicts has emerged as a significant strategic necessity for national economic and societal development[1]. Disruption in the supply chain refers to an event that interrupts the flow of goods or services, potentially resulting in severe outcomes such as financial losses, reduced operational efficiency, and decreased competitiveness[2,3]. In specific industry sectors such as semiconductors, information technology, and artificial intelligence, the presence of monopolistic enterprises is notable. Should these firms face bankruptcy, the resultant disruption in the entire supply chain could lead to unimaginable consequences for global trade supply chain.

Within semiconductor industry chain, the design, production, and processing of SoCs along with the sourcing of raw materials, are subject to stringent regulations by various countries[4,5]. In imbalanced international trade environment, vehicle manufacturers in China predominantly rely on foreign monopolistic firms for their supply of SoCs. Limited supply information exists regarding the supply connections between chip design, equipment manufacturing, and raw material supply enterprises. In cases of decoupling or supply chain disruptions, these companies face challenges in securing a suitable replacement to maintain their automotive supply continuity[2]. Hence, selecting suitable replacement cooperative firms and evaluating the effects of disruptions are of extreme concern. Knowledge graph and complex network technology provides a pivotal perspective for addressing this issue. Employing knowledge engineering to delineate the supply chain network enables the acquisition of insights into enterprises interaction, offering both a theoretical framework and practical support to bolster the robustness and resilience of supply chain[6].

However, supply information has always been the object of serious confidentiality of most enterprises, and it is extremely difficult to obtain, especially in automotive SoC industry. But there are plenty of relevant information about manufacturing and sales of automotive SoCs in open domain databases. The explosive growth of internet data contains high-value interaction information of automotive SoC supply chain. Massive amount of supply chain data recorded and collected during search engine usage and web browsing forms a comprehensive data repository. By deeply mining the hidden relationships among data, we could extract the huge supply information embedded within. Confronted with the challenge of utilizing such a diverse and extensive collection of large-scale, multi-sourced, heterogeneous data, the development of efficient methodologies for extraction of valuable insights from this data emerges as a critical issue requiring immediate attention[7].

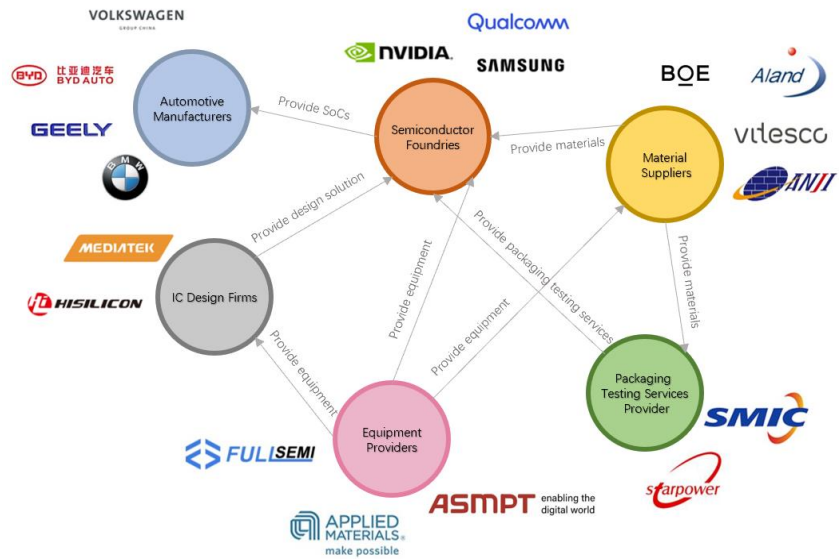

Fig. 1. Examples of interaction behaviors among firms of Chinese automotive SoC industry.

To address these challenges, we introduce an Open Supply Knowledge Extraction and Complement (OSKEC) approach. This approach is designed to extract and complete the cooperative relationships within supply chain from open domain database according a standard process. It identifies appropriate suppliers based on up-to-date open information and generates a real-world supply chain cooperative network.

This supply chain network could be regarded as a Complex Adaptive System (CAS). A CAS is defined as an "interconnected network of multiple entities that exhibits adaptive action in response to both the environment and the system of entities itself"[8–10]. To model this complexity in the CAS of automotive SoC supply chain network, two Interaction and Disruption Models (IDMs) are developed. Our simulation results indicate that automotive SoC companies can boost their resilience by enhancing their innovation and cooperation capabilities. In addition, we find it is more advantageous for an enterprise to suffer risks independently rather than transferring them to business partners under certain conditions.

All in all, this study models the interactions and disruption behaviors by constructing a network of the Chinese automotive SoC supply chain applying the OSKEC-IDM framework, which could be divided into three stages. First, OSKEC approach is proposed to extract relevant open domain knowledge. Second, data is gathered from open domain to establish a foundational firm entity dataset. The relationships between firm entities are complemented through inference technology, facilitating the expansion of scale of supply chain network. And the characteristics of the supply chain network are analyzed. Subsequently, two IDMs, based on recovery capacity (RC-IDM) and risk transfer (RT-IDM), respectively, are developed to simulate the complex behaviors of Chinese automotive SoC enterprises. we explore proactive strategies that enterprises can implement to enhance the supply chain resilience against disruptions. In Conclusion, this research proposes a practical and effective supply chain network construction framework and provides invaluable insights in Chinese automotive SoC supply chain to help enterprise strategy maker to understand the complex and dynamic interaction behaviors.

## 2. Literature Review

This section covers the literature related to the modeling of supply chain network structures and their interaction and disruption behavior as a complex adaptive system.

### 2.1 Supply chain model construction

Initially, models of the supply chain are proposed as linear and sequential, illustrating the chain as a simple series of nodes such as suppliers, manufacturers, and retailers, each connected to the next[11]. This approach often leads to oversimplification of the supply chain structure. Chain models typically assume a direct flow of products, information, and finances from one stage to the next without considering the multiple interactions that might occur at each node[12]. To overcome these limitations, modern supply chain models have evolved to incorporate network-based perspectives and advanced analytics.

Massive simulation experiments were carried out for modeling supply chain networks, which provide valuable insights into complex system behaviors under various scenarios, but with inherent limitations in the lack of real-world complexities consideration[13–17]. Additionally, these models often lack access to real-world data, significantly limiting their accuracy and reliability. These limitations stem from the simplifications and assumptions necessary for computational models and simulated data, which unable to fully capture the intricate dynamics, unpredictability, and variability of actual supply chains. This can result in a gap between the theoretical outcomes produced by simulations and the practical experiences observed in real-world operations. To address these issues above, researchers increasingly paid attention on real-world supply chain network structures. Pathak et al. discussed and described the features of American automotive supply chain network, obtained that the network environment type plays a decisive role in the formation of structures in the process evolution[8,18]. Kim Y et al. used social network analysis methods to study the linear supply chain structure of byers and suppliers, and applying this framework to the analysis of three actual automotive supply chain network case studies[19]. Zeng Y et al. introduced a method based on network load entropy to assess the vulnerability of networks and demonstrated through modeling and analysis of a cluster-type supply chain network in a specific industrial area, illustrating the effectiveness of this method in vulnerability assessment[20]. Kito T et al. have discussed the impact of market fluctuations, product standardization, technological advancements, and financial dependencies among companies on the structure of automotive supply chain networks, by analyzing actual data indices from these networks. Subsequently, they analyzed the comprehensive data in automobile parts supply chain network and clarify the closeness between strategies and product type[21,22].

More and more researcher modeling supply chain structure through real-world data, Lu and Cai et al. utilized business data from the Chinese National Important Products Traceability System (NIPTS) to construct a triple-layer supply chain model encompassing farm, slaughterhouse, and retailer stages[23,24]. Brintrup et al. assembled a large-scale empirical data set on the supply chain of airbus and apply the new science of networks to analyze how the industry is structed[22]; Calatayud et al. explored the risks that international freight flows are exposed to as a function of multiple complex structure of liner shipping networks and showed that the country plays significant role in the network[25]. Substantial research has shown that employing real-world data enables a more precise characterization of the nodes representing businesses within supply chain networks, and enhances the description of the relationships among these enterprises[26].

However, the aforementioned studies generally depend on specialized datasets or assumptions that may not accurately mirror current or future conditions, leading to a discrepancy between the theoretical supply chain structure and the actual structure observed in real-world. Therefore, while these models are useful for understanding general structure and preparing for predictable scenarios, they should be used with caution and supplemented with up-to-date real-world data whenever possible. In addition, the exploration of their structure of supply chain remains limited due to data security concerns and governmental influences in certain industries such as automotive SoCs.

**2.2 Interaction and disruption models of supply chain**

Traditional supply chain model, such as linear programming, network flow model and transportation model, typically emphasize linear and local chain structures, focused on direct upstream and downstream relationships, which thus fall short in capturing the complexity of modern supply chains and fail to adequately represent the multiple attributes of supply chain nodes and the diverse interactions across different levels of the supply chain network[12,27]. Furthermore, with the increasingly complexity of supply chain network, the dynamics of interaction within the supply chain network are critical for its effective operation and resilience.

The interaction and disruption behaviors of supply chain network are increasingly being analyzed from the perspective of complex adaptive systems. [2,28]. Based on the seminal work of Choi et al., a CAS is a self-organizing system, and it reconfigures its internal and external linkages to continually evolve over time[10]. Subsequently, Kim et al. and Nair, Narasimhan note that complex adaptive system is a useful theory in describing supply chain network structures[29,30]. Pathak et al. term supply networks as a typical case of CAS because a supply chain will adapt via interactions of nodes within the network and evolve over time [16]. These interactions are not merely transactional but involve strategic collaboration, information sharing, and mutual dependency among various stakeholders, including suppliers, manufacturers, and customers.

However, the interconnected nature of modern supply chains makes them susceptible to a wide range of disruptions, which can propagate quickly across the network, affecting multiple entities simultaneously[22,25,31]. The disruption of supply chain refers to unexpected event or series of events that significantly interrupts the normal flow of goods and services within a supply chain network, potentially leading to delays, increased costs, reduced quality, or complete failure in meeting customer demands till bankruptcy[2,3,32,33]. Disruptions in the supply chain could arise from a variety of sources, including natural disasters, geopolitical tensions, economic fluctuations, and technological failures[31,33].

Viewing the supply chain network as a CAS provides a valuable approach for studying the effects of disruptions. In the disruption process, a company's behavior is shaped by its specific characterize, making it intriguing to investigate how firm's business interaction strategies evolve. For example, Kim et al. find that if a single central node acts as the focal point for aggregating supply and distributing demand, any disruption to this critical node could potentially lead to the complete cessation of operations across the entire supply chain network[29]. Wang et al. investigate the financial repercussions at the firm level resulting from supply chain disruptions during COVID-19. They explore how companies' supply chain diversification strategies—encompassing diversified suppliers, customers, and products—mitigate the adverse impact on firm performance[33]; Takawira et al. explore the key lesson that a comprehensive demand management strategy need to be adopted by pharmaceutical companies to improve demand visibility, responsiveness and supply chain resilience[34]. The impact of company strategies in the field of disruption can be profound, leading to significant market shifts and resilience reduction within the industry supply chain. Moreover, the ripple effects of strategies can extend beyond the immediate supply chain, affecting economies and societies at large. Thus, understanding the mechanisms of interaction and disruption within supply chains is paramount for developing effective strategies to mitigate risks and enhance resilience. IDM facilitates a comprehensive understanding of the repercussions of in complex adaptive systems, which is an effective methodology to simulate the effects of disruption spread within sociology[35], ecology[36] and medicine[37], and enhances the capability for improved management and strategic planning across diverse sectors, enabling stakeholders to address and mitigate potential challenges effectively.

The discussion above illustrates the importance of developing appropriate models to capture the interaction and disruption behaviors within supply chain CAS. However, previous models of supply chain networks lack a detailed exploration of the mechanisms of disruption impacts and the global changes across the entire network. This highlights the need for IDMs that can provide a more comprehensive understanding of how disruptions affect these complex systems.

## 3. Research Methodology

### 3.1 Overall Framework

This study aims to understand the complex adaptive interactions and disruption behaviors in supply chain network. For this purpose, we propose an OSKEC-IDM framework for supply chain network construction and cascade failure stimulation. As shown in Fig 1, this framework consists of two modules, the open supplier knowledge extraction and complement (OSKEC) approach for supply chain entities recognition, name disambiguation and relationship inference, and the interaction and disruption model (IDM) for cascade failure simulation in supply chain network.

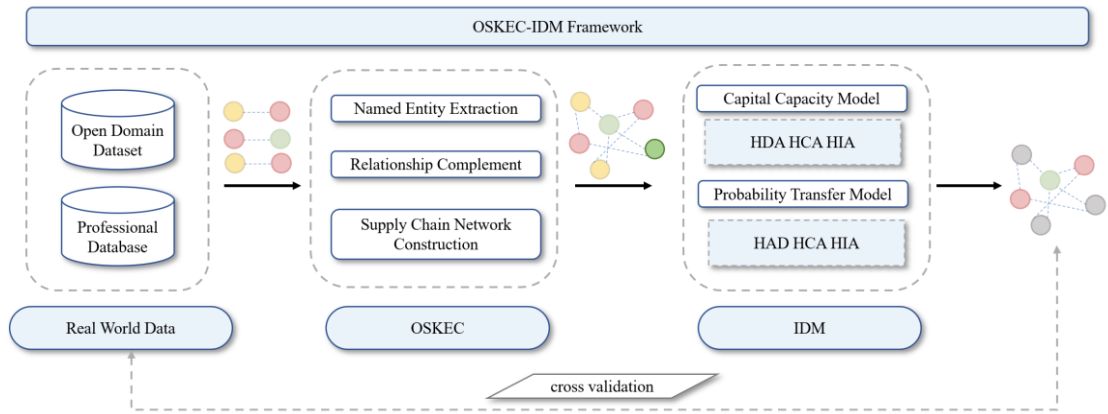

Fig. 2. Architecture of the OSKEC-IDM framework.

Initial data collection was achieved by scraping a database to gather information on firms, encompassing both their financial data and the relational data among companies. Besides, we get firm entities from open domain dataset mainly including relevant research reports and news reports. Next, the relationships among suppliers are deduced using LLM (Large Language Model) and Knowledge Graph (KG) inferencing technology, which helps expanding the scale of supply chain network. Subsequently, the characteristics of the network are analyzed, and models are employed to assess the impact of disruptions in various scenarios. Finally, we explore proactive strategies that companies can implement to enhance their supply chain resilience against disruptions.

In OSKEC-IDM framework, the input is both open domain data and professional data specific field, which includes comprehensive enterprises cooperation corpus and the output is simulation result of disruptive networks. Besides, dynamic cross validation is used to compare the simulation results with real-world data, ensuring that OSKEC-IDM framework accurately reflects real-world scenarios.

### 3.2 The Open Supplier Knowledge Extraction and Complement Approach

As a core component in new energy vehicles, automotive SoC is essential for integrating multiple functionalities and systems and ensuring high performance and reliability in the automotive industry. The development and operation of automotive SoC require the collaboration and supply from numerous chip design and manufacturing firms, which generates massive relationships and interactions among different firms. Simplistically, supplying relationships can be streamlined into a series of triplets like <entity *a*, relationship, entity *b*> within a supply chain network. For example, the triplet <Material suppliers, provide materials, Semiconductor foundries> in Fig. 1 represents that the material suppliers provide materials to semiconductor foundries.

Given the challenge of accessing comprehensive supply information on automotive SoC field, we propose an open supplier knowledge extraction and complement (OSKEC) approach for network construction of suppliers, as shown in Fig. 3. Open domain data refers to open resource dataset and Internet website information that are freely accessible by the public without any copyright or proprietary restrictions. This type of data is crucial for research, development, and educational purposes, as it promotes transparency, innovation, and collaboration across various fields. In contrast, professional data often requires specialized technical methods or incurs commercial costs to gather, typically involving proprietary or restricted datasets that are not openly available for public use.

In the OSKEC approach, a method tailored for low-resource environments to construct supply chain networks is developed. First, for open domain data, a substantial array of firm entities named as "Open Entities" is extracted from diverse industrial field. Subsequently, cross-domain transfer learning is used via GPT-NER to identify the firm entities. Concurrently, entity and relation extraction from professional database is proceeding, incorporating open entities to "Basic Entities". Following this integration, the preprocessing of firm entities is executed using fuzzy matching technologies to rectify discrepancies in name conventions, exemplified by variations such as "NVIDIA," "NVDA," and "Nvidia Corporation.". Ultimately, relation completion is carried out in the matched triplets leveraging Large Language Model (LLM) and Knowledge Graph (KG). This method relies on supply relationships sourced from professional databases, which ensures a feasible strategy for construction of supply chain network even when definite supply data is lacking.

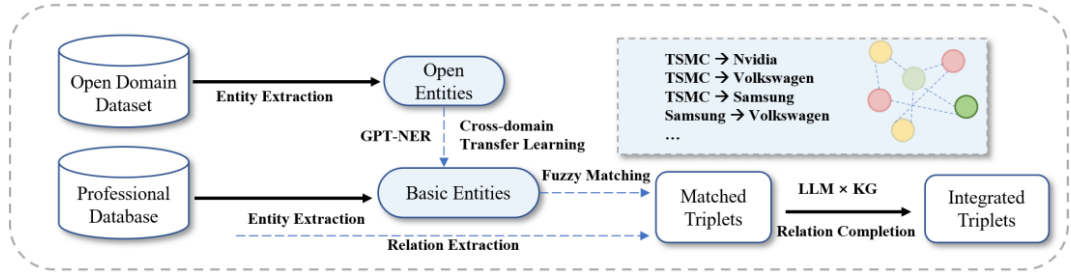

Fig. 3. The OSKEC approach.

### 3.2.1 Entity Extraction

Open domain includes massive supplier information. Named entity recognition (NER) is used to find firm entities in supply chain from open domain datasets and professional databases. we could extract various open entities using NER technology. The entities are not all relevant to the specific industry field we want to explore, but they are important resource to solve the low-resource problem of domain entity recognition like automotive SoC suppliers.

NER is a typical sequence labeling task, give a sentence $X = \{x_1,...,x_n\}$, it assigns an entity type $y \in Y$ to each word $x$, where $Y$ denotes the set of entity labels and $n$ denotes the length of the given sentence. A common approach to is to formulate it as a sequence labeling task. Subsequently, fine-tunning GPT-NER is used to resolve the problem that there is the gap between the two tasks the NER and LLM prompt construction, the former is a sequence labeling task in nature while the latter is a text-generation model. GPT-NER exhibits a greater ability in the lower-source and few-shot setups, when the amount of training data is extremely scarce, GPT-NER performs significantly better than supervised models, which is suitable for the domain of industry field lacking numerous supply information. GPT-NER can be decomposed into the following three steps: firm datastore construction, entity representation extraction and few-shot demonstrations for LLM prompt. As shown in Fig. 4.

(1) Firm datastore construction

The train dataset comes from open domain in Specific SoC supply chain, we use automated scripts to label approximately 8,505 train records via "BIO" label method[38], then the NER model is fine-tuning by this dataset.

(2) Entity representation extraction

In order to identify firm entities in specific field, we adapt fine-tuning pre-trained models for NER task. This process involves fine-tuning the model to align with specific tasks and domains. adapting them to specific tasks and domains. By doing so, we enhance the models' performance in recognizing entities within the specialized domain.

Representation extraction aims to obtain the high-dimensional vector representation for each token within the input textual sequence. We use BERT[39] as the encoder model to represent the high-dimensional vector of corpus related to supply chain enterprises, and the output of BERT is $h_i \in \mathbb{R}^{m \times 1}$, where $n$ denotes the length of the input sentence and $m$ denotes a variable parameter of the dimension of the vector. Then each embedded high-dimensional vector $h$ is sent to a multi-layer perceptron and then generates the distribution over the named entity vocabulary using the softmax function:

$$p_{\text{NER}} = \text{softmax MLP}(h \in \mathbb{R}^{m \times 1}) \tag{1}$$

Based on $p_{\text{NER}}$, the embedded high-dimensional vectors are classified into labels according to a softmax layer, where "Nvidia" and "wafer" are two recognized entities.

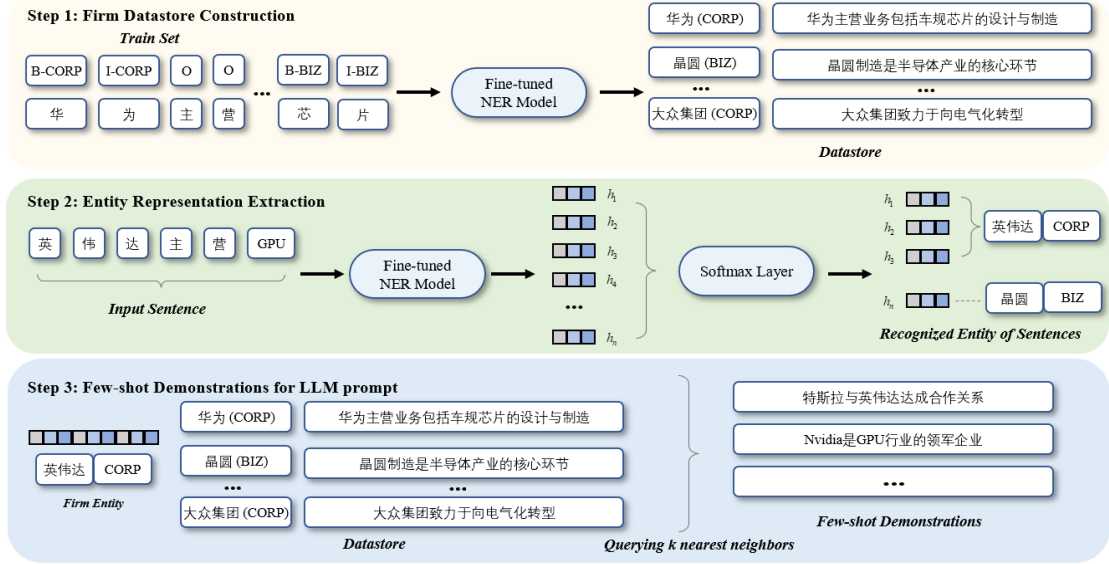

Fig. 4. An example of entity-level embedding and few-shot demonstrations in GPT-NER.

(3) Few-shot demonstrations for LLM prompt

In step 3, the few-shot demonstration is appended to the prompt. It serves as the following two purposes: First, it regulates the format of the LLM outputs for each test input, as LLMs will (very likely) generate outputs that mimic the format of demonstrations. This is vital for the NER task as we need the output format to be consistent so that we can parse the output in the form of natural language to NER results; Second, it provides the LLM with direct evidence about the task and references to make predictions.

We get firm entity vector representation from step 2, then we could retrieve $k$ nearest neighbor (kNN) of the input sequence from the training set[40] via cosine similarity. It is worth mentioning that we use entity-level representations instead of sentence-level representations. The shortcoming of kNN based on sentence level representations is obvious: NER is a token level task that focuses more on local evidence rather than a sentence-level task, which is concerned with sentence-level semantics. A retrieved sentence (e.g., It is a semiconductor company) that is semantically similar to the input (e.g., Nvidia is a semiconductor company) might shed no light on the NER the input contains. In the example above, the retrieved sentence contains no NER and thus provides no evidence for tagging the input.

To resolve the issue above, we need to retrieve kNN examples based on entity level representations rather than sentence level representations. For a given input sequence with $N$ entities, we first iterate over all tokens within the sequence to find kNNs for each entity, obtaining $K \times N$ retrieved tokens. Next, we select the top $k$ tokens from the $K \times N$ retrieved tokens, and use their associated sentences as demonstrations.

The demonstration sequentially packs a list of examples, where each example consists of both the input

sequence *X* and the output sequence *W*:

*Input: [Example Sentence]₁*

*Output: [Labeled Sentence]₁*

*...*

*Input: [Example Sentence]ₖ*

*Output: [Labeled Sentence]ₖ*

where *k* denotes the number of demonstrations.

**3.2.2 Cross-domain Transfer Learning and Fuzzy Matching**

Supplier information is often treated with strict confidentiality by various enterprises, making it exceedingly difficult to access, particularly in the specific industry like automotive SoC industry. In order to overcome the difficulties and compensate for the lack of data, cross-domain transfer training has been utilized in NER tasks. Integrating external knowledge sources, such as financial databases, stock information, and marketing data, could further improve NER performance by supplying extra context for named entities not covered in the training dataset. Blind mixing of all the available datasets has been proved not to be desirable[41], thus we select open domain related to automotive and IC industry, it's obviously performed better that some irrelevant data source. Besides, numerous studies have shown that GPT-NER is based on fine-tuning pre-trained models and performance better in few shot tasks[38,42,43].

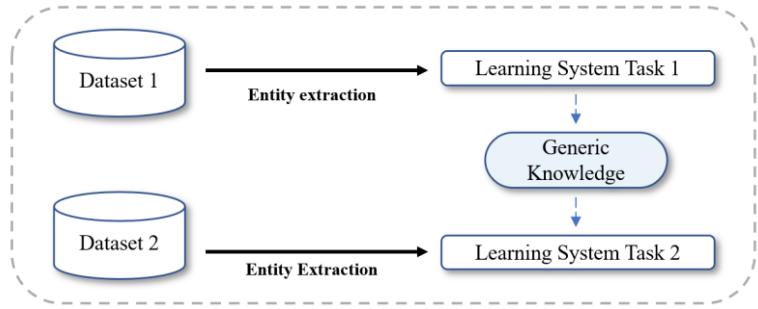

Fig.5 Cross-domain transfer learning.

In OSKEC, this approach emphasizes the preservation and application of insights acquired from addressing a particular issue to supply chain network construction. It involves educating a model using extensive datasets, then leveraging this acquired knowledge to benefit another model trained with limited data in Chinese automotive SoC industry. By utilizing information across various domains, this method significantly improves entity classification in languages with scarce resources. However, transfer learning introduces the challenge where a single firm entity may have different names across various fields. To address this issue, fuzzy matching techniques are employed.

We use Levenshtein distance, also known as the edit distance, to measure the difference between pre-processed similar firm entities. The Levenshtein distance (represented as *L*) between two firm strings *a* and *b* can be calculated using a dynamic programming approach. Then, the similarity *S* of these two firm strings can be represented as

$$S = \left(1 - \frac{L}{\max(|a|,|b|)}\right) \quad (2)$$

where $|a|$ and $|b|$ represent the length of firm strings *a* and *b*, respectively. For Example, we calculate the *S* between NVIDA and NVDA, and get the *S* as 0.89. The higher the similarity score between firm entities is, the more similar the two entities are. Then we make all firms with a similarity score above 0.6 are grouped as the same firm. Next, we concentrate on supply relation completion between different companies.

**3.2.3 Relation Completion**

Triplets can naturally be represented as a knowledge graph (KG). In automotive SoC supply chain relation completion, the supply knowledge graph G is a directed graph, in which vertices represent entities $\varepsilon$, and each edge can be represented as a triplet $(h,r,t)$, where $h$, $r$, and $t$ correspond to the head entity, relation, and tail entity, respectively. Relation completion task is to infer missing triplets given an incomplete knowledge graph $G$. Under the commonly used entity ranking evaluation method, tail entity prediction $(h,r,?)$ requires ranking all entities given $h$ and $r$, and head entity prediction $(?,r,t)$ is similar. In this study, for each triplet $(h,r,t)$, we add an inverse triplet $(t,r^{-1},t)$, where $r^{-1}$ is the inverse relation of $r$, thereby only needing to deal with the tail entity prediction problem[44]. In this section, the purpose is to complete matched triples through the collaboration of LLM and KG.

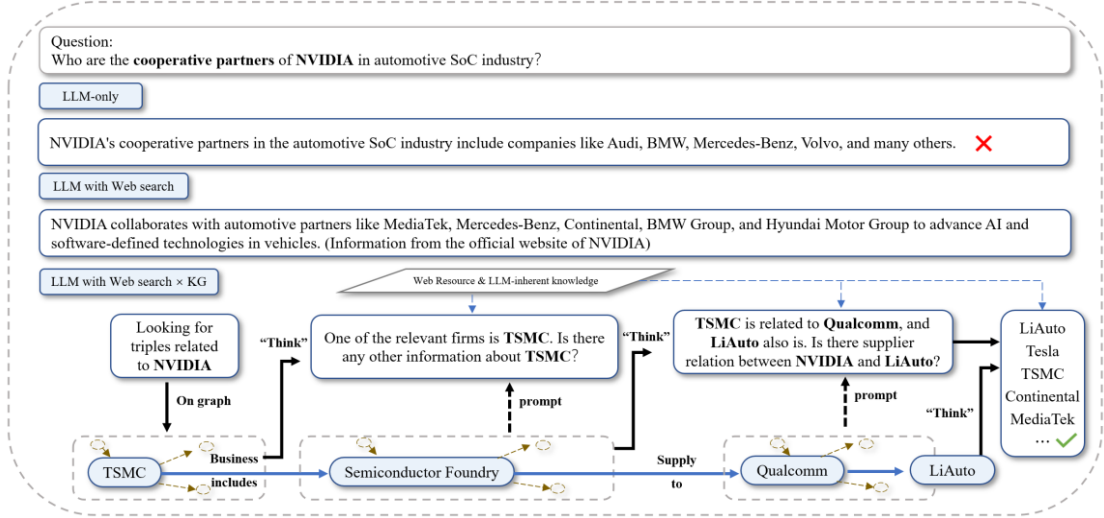

Fig.6 An example of relation completion using LLM×KG.

LLMs often demonstrate certain limitations in their ability to absorb new information, in generating hallucinations, and in the transparency of their decision-making processes[45]. We integrate specific supply chain KG into LLMs, thereby expanding our acquired triplets. The steps are as followed: (1) Traversing representative firm entities in acquired triplets and integrating them into the prompt; (2) Looking for entities related to the representative firm entities and integrating into the prompt, then LLMs generate answer with our prompt, web resource and its inherent knowledge; (3) Continuing step (2) until obtain enough firms which have supply relationship with the representative firm entities.

### 3.3 Supply Chain Network Construction and Analysis

It's natural to convert the triplets in KG to a supply chain network. The triplets of nodes and relationships can be directly mapped to the nodes and relationships in a complex network model. We first analyzed the basic topological structure of this supply chain network. In addition to conventional network metrics, we also obtained the registered capital information for each company from the professional database, which serves as an indicator of their influence within the supply chain. Then we calculated various centrality metrics, and compared their correlations. Furthermore, as previously mentioned, there are plenty of studies view supply chains as complex adaptive systems for analysis[10,16,29,30]. Therefore, we constructed two complex adaptive models to explore the propagation of supply chain disruptions across the network and their overall impact on the supply chain network.

#### 3.3.1 Network Construction

OSKEC approach receives automotive SoC supply chain knowledge and outputs integrated triples. The triplets in knowledge are primitively to be transformed into edges and nodes within complex network. For a triplet $(h,r,t)$, the head entity and tail entity can be represented as firm nodes $v_i$ and $v_j$, and the $r$ is represented as supply relation $e_{ij}$.

There is normally bi-directed cooperative relation between two firms in Chinese automotive SoC industry. Thus, we use an undirected network to model the supply relation between two automotive SoC firms to address this issue. In this network, $v_i$ and $v_j$ denotes the firm node $i$ and $j$ of automotive SoC firms, respectively. $e_{ij} = e_{ji} = 1$ denotes that there is supply relationship between firm $i$ and firm $j$, otherwise there is no supply relation in them.

Maintaining only a single fundamental supply relationship between two supplier entities (for example, supply relationship between TSMC and Nvidia). Furthermore, the automotive SoC supply chain is represented as an undirected network denoted by $G(V,E)$.

### 3.3.2 Network Analysis

Understanding complex network indices and metrics how corresponds real-word complex trade phenomenon is a challenging task. However, we still could get some insights from these complex indices owing to they could represents complex structure in Supply chain network in a way. For example, the average path length in networks denotes the average number of steps along the shortest paths for all possible node pairs, and the shorter average path lengths typically suggest a more efficient and responsive supply chain. Moreover, a smaller network diameter might indicate a compact network where suppliers and manufacturers are closely linked, potentially reducing transport times and costs but perhaps increasing vulnerability to localized disruptions. From community partition perspective, modularity is used to define and quantify the cooperative relationships within supply chain network. Modularity of Louvain algorithm is defined as:

$$Q = \frac{1}{2m}\sum_{ij}(S_{ij} - \frac{k_i k_j}{2m}) \cdot \delta(C_i, C_j) \tag{3}$$

Where $m$ denotes the total number of edges within the network, $S_{ij}$ indicates whether there is an edge between node $i$ and node $j$, with 1 for an edge present and 0 otherwise. Furthermore, $K_i$ represents the aggregate count of edges connected to node $i$. $C_i$ and $C_j$ represent the community affiliations of nodes $i$ and $j$ belong, respectively. In instances where both nodes are members of the identical community, $\delta(C_i, C_j)$ is assigned a value of 1; conversely, it is set to 0.

Further, we compare a set of classic network centrality metrics in 4.3 and they are defined as follows in supply chain network. Firm nodes that hold the higher page rank/hub value within a supply chain normally are key influencers or critical suppliers. These nodes are pivotal in maintaining the flow of goods and information across the network. Clustering coefficient and triangles are both used to describe the local connection features in a supply chain network, which suggest the degree that companies tend to cluster around key players. Nodes with high eigen centrality in a supply chain are those that are not only well-connected but also connected to other well-connected nodes. As for closeness, firm nodes with high closeness centrality in a supply chain means they could quickly interact with all other nodes, making them efficient for roles like centralized manufactures or distribution centers that need to send or receive goods to (from) all parts of the network swiftly. High between centrality suggests firm nodes that frequently occur on the shortest paths between other nodes in the supply chain, hold strategic importance as control points for the flow of goods and information. Enhancing these firm nodes can significantly impact the supply chain's robustness and resilience.

### 3.4 The Interaction and Disruption Model

In this section, we introduce two interaction and disruption models based on recovery capacity and risk transfer, respectively, from complex adaptive system perspective.

### 3.4.1 Interaction Behavior

In supply chain network, undirected edges represent bilateral cooperation between firm nodes, including upstream and downstream collaboration. Hence, any business action of one will influence all its connected firm nodes, which also means that the disruption will spread along with the bilateral cooperation till the whole Chinese automotive SoC supply chain network is completely destroyed. Besides, due to the great disparities in registered capital among enterprises in real-world Chinese automotive SoC supply chain, we take the logarithm of registered capital to mitigate this impact. This approach helps balance the differences in capital magnitude among firms and more accurately reflects their importance in the supply chain. Both the interaction behaviors and disruption methods in this research are based on the assumptions above.

Each enterprise seeks to find am optimal strategy to ensure itself in a safe position in the supply chain under this propagation of disruption. We assumed that the risk propagation in the supply chain is sequential, that is, the state of an enterprise at a certain moment is determined by the state of its neighbor nodes and the state of the enterprise itself at the previous moment. Thus, the state of node $i$ at time of $t$ is $s_i(t)$, and the influence of its neighbor node $j$ is $\beta_j * s_j(t-\tau)$, $\beta_j$ denotes the effect of the neighbor node $j$ on node $i$ and $0 \leq \beta_j \leq 1$.

Moreover, a number of researches have proved that the more registered capital a firm has, the stronger its ability to confront to risk[31,46]. Thus, we definite the anti-risk ability of a firm is $\mu * c_i(t-\tau)$, which $c_i(t-\tau)$ denotes the logarithm of registered capital of firm node $i$ at time of $t-\tau$, $\mu$ denotes the anti-risk coefficient and $0 < \mu \leq 1$.

Furthermore, the disruptive enterprise will dispersal its risk to its neighbor nodes, which will cause a cascade failure to the whole supply chain network. On the one hand, an enterprise may opt to transfer portions of the risk from itself to its neighbors. However, once the number of its alive neighboring nodes decreases, the enterprise itself will have to undertake a greater share of the risk, which also means that the enterprise is more possible to be failure. On the other hand, an enterprise could choose to undertake all the risk independently, which may increase the likelihood of disruption. In this situation, all firm nodes will continue this process until they are disruptive. In conclusion, each enterprise relies on its supplier and forms a community with shared interests. Nonetheless, an enterprise must decide whether to offload risk to its suppliers. If an enterprise opts to transfer the risk, the immediate outcome may be favorable, enhancing its survival prospects. Besides, it will also face heightened vulnerability if its neighbor suppliers sustain damage. Selecting the appropriate strategy is crucial. Our research models the risk propagation process and assesses whether it is advisable for major firms to transfer risk to their suppliers.

### 3.4.2 Disruption Method

The largest connected component of network (also as largest connected subgraph) always is used to represent the vital features of specific social, finical, technological network[47]. We also use the largest connected component to represent the whole Chinese SoC supply chain features and ignore some boundary firm (with on edges connected with other firms).

We use Louvain algorithm to partition the communities within Chinese automotive SoC supply chain network, and there is a clear demonstration of community distribution (shown in Figure 7), where enterprises with significant influence in automotive SoC field often play a crucial role in the network. As the leaders of the supply chain, they have a significant impact on the entire supply chain's operations. By analyzing and studying the behaviors of these key companies, we can gain insights into the overall pulse of the automotive SoC supply chain. This understanding paves the way for further analysis and research on the interactions and disruption models within the entire automotive intelligent chip supply chain network. In the real world, some firms have the certain possibility to be attacked under random events, such as policies change, finical crisis and natural disasters. Under different attack policies, the cascading interruption process of the entire network is different. Choosing degree, closeness centrality and registered capital as the measures to select the initial disruptive firm nodes, and three network attack strategies are as followed:

- High degree attack (HDA): targeting nodes with the highest number of connections to quickly dismantle the network's most interconnected components.
- High closeness centrality attack (HCA): focusing on nodes that have minimize distances to all other nodes

due to their strategic positions, thereby probably disrupting the network structure fast.
- High importance attack (HIA): selecting nodes based on their risk-taking capacity (registered capital), aiming to maximize the impact of their removal.

Nodes in the network with a higher degree of connections typically hold significant control over the entire supply chain due to their extensive collaborations with other enterprises. Nodes with greater closeness centrality are closer to other nodes, indicating strategic positioning within the network. Additionally, nodes that represent influential companies with substantial capital capacities can have a profound impact on the supply chain if they go bankrupt. This research selects initial businesses for disruption based on different attack strategies, which helps us better understand how different types of disruptions affect the automotive SoC supply chain during cascading events. Based on these various network attack strategies, we employ different disruption methods to simulate how risk propagates in the automotive SoC supply chain. We consider the risk of cascading failures in the automotive SoC supply chain under the influence of multiple external coupling factors and have designed the following two supply chain disruption risk models:

(1) RC-IDM

In the process of disruption spread, the enterprise will locate between state of operation and disruption, so that we define the state of enterprise range from 1 and 0, 1 denotes that the firm is absolutely disruptive and 0 denotes the firm is operating normally. In real world, it's utterly lethal for firms when their capacity is decreased into 50% than initial capacity. Thus, we set 0.5 as the threshold value to and observe the global changes in automotive SoC supply chain network.

We definite the formula to update the state of SoC firms:

$$s_i(t) = s_i(t-\tau) + \delta(\lambda(1-s_i(t-\tau))\sum \beta_j s_j(t-\tau) - \mu s_i(t-\tau)) \tag{4}$$

where $s_i(t)$ denotes the state of firm $i$ and it ranges from 0 to 1, $\lambda$ denotes the influenced coefficient of all neighbor firm nodes, and $\beta$ denotes the influence from single neighborhood, and the $\mu$ denotes the recover coefficient of firm node $i$, $\tau$ denotes the delay operator, normally is 1, $\delta$ denotes time gap.

Pseudocode of RC-IDM

**1** Choosing $n*p$ nodes to failure under specific network attack strategy, and signed $s_i = 1, i \in (1,2,\ldots,n*p)$, $n_j \in n_i$' neighbor nodes
**2** For each time step:
**3** For node $n_i$ in SoC supply chain network:
**4**   if $s_i != 1$:
**5**     $s_i(t) = s_i(t-\tau) + \delta(\lambda(1-s_i(t-\tau))\sum \beta_j s_j(t-\tau) - \mu s_i(t-\tau))$
**6**   else:
       $s_i(t) = s_i(t-1)$
**7** Calculate $\dfrac{\sum s_i}{n}, (s_i > 0.5)$ at each time step as affected ratio
**9** Return affected ratios and timesteps

(2) RT-IDM

In real world, the enterprises normally have two strategies: 1) transfer strategy: transferring the risk to its neighbor firm nodes; 2) absorb strategy: undertaking the risk by their self. It will be alive in short time once the firm choose transfer strategy, however, the disruption risk will be increased rapidly in next time step because the possibility of the firm is closely relevant to its neighbor firm nodes.

The disruption possibility of the enterprise is defined as:

$$p_{failure} = \frac{\sum s_j(t-\tau)}{\sum_{if\ s_j(t-\tau)=1} s_j(t-\tau) + \sum_{if\ s_j(t-\tau)=0}(s_j(t-\tau)+1)} \tag{5}$$

where $s_j(t-\tau)$ denotes the firm node's state at time of $t-\tau$.

Pseudocode of RT-IDM

---

**1** Choosing $n_l$ nodes to failure under specific network attack strategy, and signed $s_i = 1, i \in (1,2,\ldots,n_l)$, $n_j \in n_i$' neighbor nodes

**2** For each time steps:

**3**   For node $n_i$ in SoC supply chain network:

**4**     if $s_i != 1$:

**5**       if strategy 1:

**6**         $p_{failure} = \dfrac{\sum\limits_{if\ s_j(t-\tau)=1} s_j(t-\tau)}{\sum\limits_{if\ s_j(t-\tau)=1} s_j(t-\tau) + \sum\limits_{if\ s_j(t-\tau)=0}(s_j(t-\tau)+1)}$

**7**       judge the state of node $n_i$ based on $p_{failure}$

**8**       if $s_i(t) = 1$:

**9**         $c_i(t) = max(c_\triangle, c_i(t-\tau) - \triangle c)$

**10**         for node $n_j$:

**11**           $c_j(t) = max(c_\triangle, c_j(t-\tau) - \beta_j \triangle c)$

**12** Calculate $\dfrac{\sum s_i}{n}, (s_i = 1)$ at each time step as affected ratio

**13** Return affected ratios and timesteps

---

If the firm node chooses transfer strategy, its capacity will be reset as:

$$c_i(t) = max(c_\triangle, c_i(t-\tau) - \triangle c) \qquad (6)$$

If the firm node chooses absorb strategy, the capacity of its neighbor firm node $j$ will be reset as:

$$c_j(t) = max(c_\triangle, c_j(t-\tau) - \beta_j \triangle c) \qquad (7)$$

where $c_i(t)$ denotes the capacity of firm node $n_i$ at time of $t$, and $c_\triangle$ denotes the minimum capacity, $\beta_j$ denotes the attenuation coefficient for neighbor firm node $n_j$.

## 4. Empirical Results and Analysis

In this section, we take the SoC supply chain in China as an example to illustrate the application of OSKEC-IDM framework in construction, typological analysis, and disruption simulation of supply chain network.

### 4.1 Supply Chain Data Sources

Open domain includes numerous relevant text information about automotive SoC industry and we collect open domain data from professional research reports and websites using a data scraper. The primary sources we targeted include two leading websites within the Chinese automotive industry : (1) **https://www.autohome.com.cn** and (2) **https://auto.gasgoo.com**. Regarding the professional database, we use **CSMAR (**China Stock Market & Accounting Research) and **Wind** as the supply chain data source. CSMAR database provides comprehensive and reliable data of top 5 suppliers and clientele pertaining to Chinese listed companies in Chinese SoC industry from 2002 to 2020, and Wind database provides customer and supplier data disclosed by more than 20,000 companies in the A-share and Hong Kong stock markets, then we use the data related Chinese automotive SoC from 2000 to 2020.

### 4.2 Network Construction

The OSKEC approach combines open domain information with professional databases in Chinese automotive SoC industry, confronting the challenge of limited access to confidential supplier information. In the end, it identifies 7,784 firm entities and 8,576 triplets within the Chinese automotive SoC industry. Each triplet represents a cooperative relationship between two firm entities, which lays a solid foundation for supply chain network construction and analysis.

The evolution of triplets is shown in Table 1, which provides a summary of the basic statistics of the supply networks including number of nodes, number of edges, characteristic path length (the average shortest path length between a pair of nodes), diameters (the maximum of shortest path length between any two nodes), and modularity (the metric to measure the cluster degree of node communities).

In investigating the dynamics of interaction and disruption within the automotive SoC supply chain, focusing on the network's largest connected component effectively isolates and disregards solitary nodes—suppliers or entities characterized by limited interconnections within the network. Consequently, it accentuates the supply chain's nucleus, which is paramount for supply chain operational dynamic analysis. Moreover, it simplifies the supply chain's complexity, facilitating the identification of opportunities for enhancement, risk reduction, and optimization[46,48]. Therefore, supply chain network within the subsequent discussion denotes the largest connected sub-network identified within the integrated triplet network, which contains 1,369 nodes and 2,241 edges.

Tab 1. Basic network statistics.

|  | Number of nodes | Number of edges | Average Path length | Diameter | Modularity |
| --- | --- | --- | --- | --- | --- |
| Basic triples | 6,579 | 1,995 | 5.70 | 17 | 0.72 |
| Matched triplets | 6,469 | 1,983 | 5.37 | 15 | 0.68 |
| Integrated triples | 7,736 | 2,476 | 5.36 | 14 | 0.71 |

Visualization of the supply chain network is shown in figure 5. The size of a node is proportional to the corresponding firm's size, measured by degree of node, and the color of a node represents the network cluster (generated with Louvain cluster algorithm[49]) the node belongs to. Utilizing the Louvain algorithm, each firm within the automotive SoC supply chain network is allocated to its corresponding community. This method facilitates the identification of modular structures within the supply chain, enhancing our understanding of the intricate relationships and interdependencies among firms in the automotive SoC ecosystem.

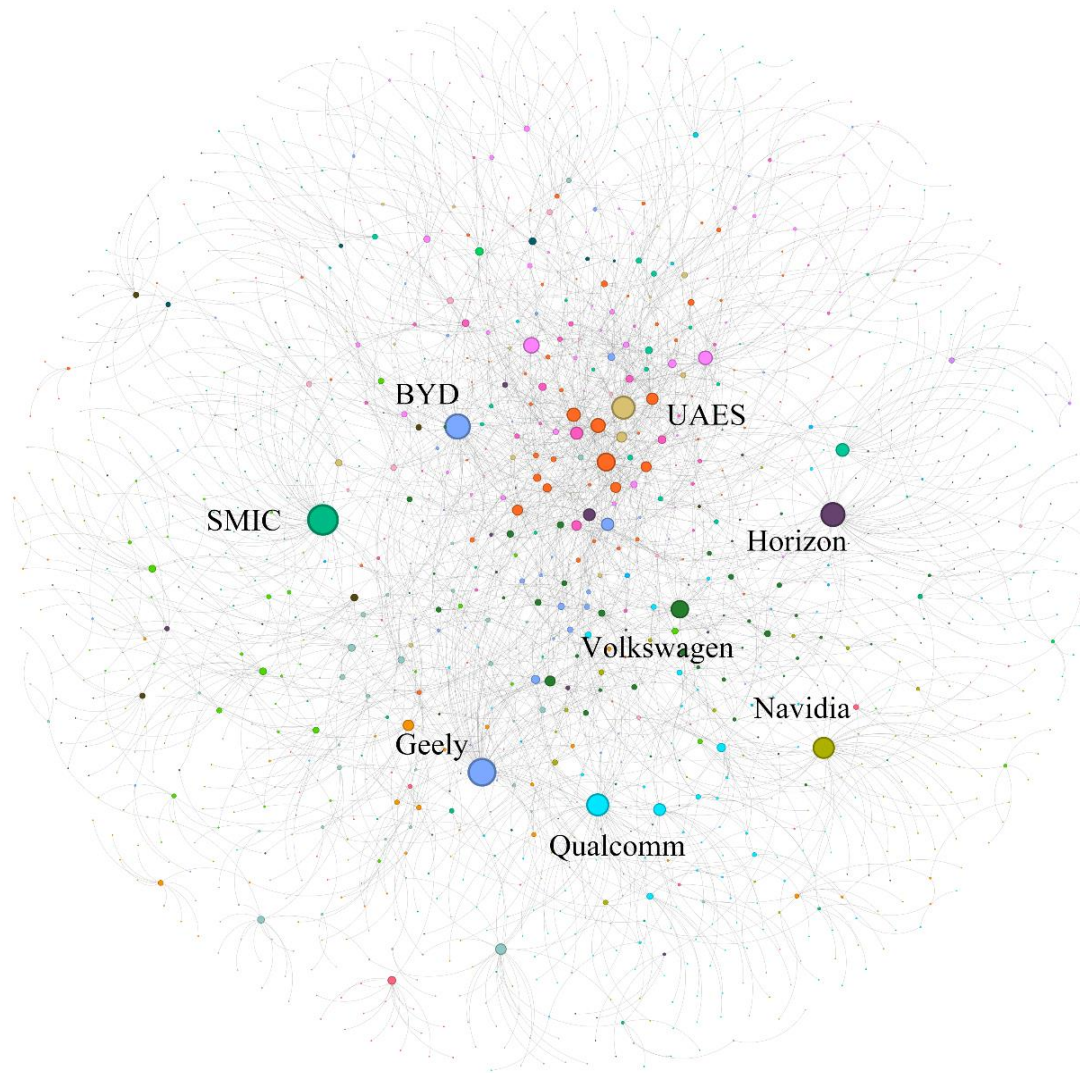

Fig.7 Automotive SoC supply chain network in China market.

**4.3 Topological Analyses**

Numerous networks exhibit scale-free characteristics, including global trade network[50], social networks[51], human disease network[52] and the Internet[53]. We also discovery that the automotive SoC supply chain network in Chinese market also exhibits these scale-free features with complex typologies. A small number of firms with large degrees dominate, while numerous nodes have a small degree, indicating that certain core firms hold significant positions in the automotive SoC supply chain. In other words, most nodes in the network have few neighbors, while there are few nodes with many neighbors. We also observe a similar trend in registered capital distribution when plotted on a log scale of frequency, with the unit of capital being ten thousand. This clearly indicates the presence of some firms in a monopolistic position, around which numerous other firms are clustered within the Chinese automotive SoC supply chain.

Previous studies showed a supply network with scale-free distribution is usually robust against random failures but is more fragile when important high degrees are attacked[2,14]. We will evaluate this later in our simulation analysis.

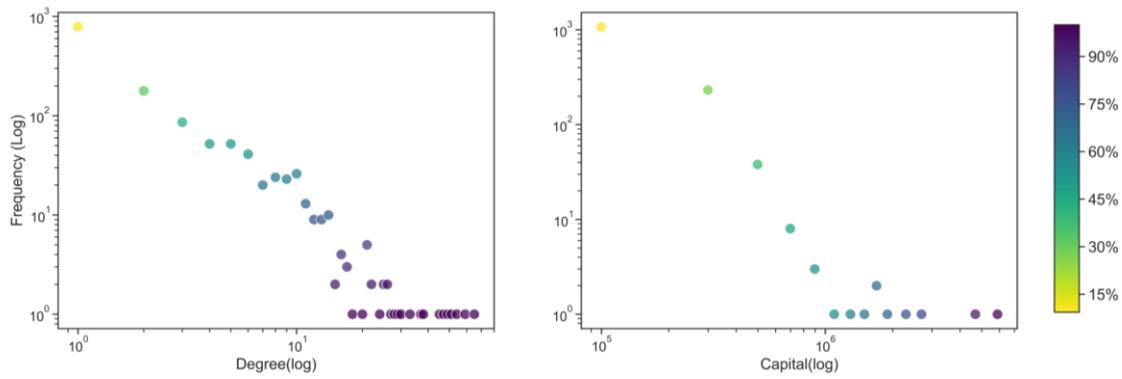

Fig.8 Degree and capital distribution.

In the analysis of automotive SoC supply chain network, registered capital of an enterprise is as a measure of its influence within the network in the real world. By analyzing the correlation between a firm's degree of connections and its registered capital, we have identified that the correlation between these factors is extremely weak. This observation underscores the differentiation between capital and connectivity as indicators of influence within the supply chain: while capital denotes tangible influence, the degree of connections underscores a firm's strategic positioning within the network's architecture. As shown in figure 9.

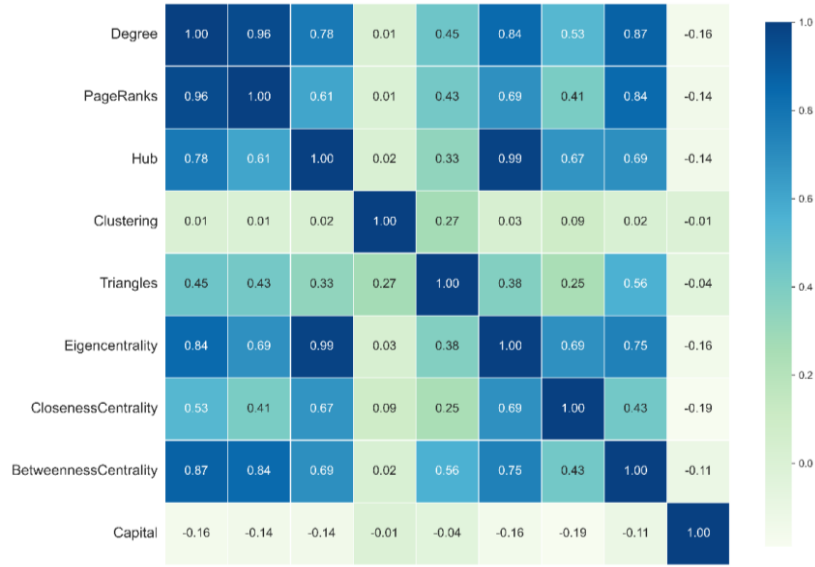

Fig.9 Correlation of network features.

Furthermore, our investigation sheds light on intriguing phenomena within the network's topological structure. Notably, nodes representing firms with extensive connections also exhibit high betweenness centrality and PageRank values. Betweenness centrality quantifies the extent to which a node presides over the shortest paths traversing other nodes, essentially gauging its role as a conduit or linkage across disparate sections of the network. The PageRank algorithm, a seminal contribution by Google, assesses the significance of nodes within the World Wide Web[54]. Moreover, a strong relationship indicating that nodes serving as hubs (connecting to many high-quality nodes) also have high eigen centrality scores, reinforcing their central role.

The intertwined nature of metrics such as Degree, PageRanks, BetweennessCentrality, and Eigencentrality intimates that the most central nodes wield considerable influence over the interaction behavior and disruption process. These central nodes emerge as vital cogs within the supply chian network, indispensable to the supply chain's operational efficiency and resilience. The observed positive correlations between local clustering indicators (Clustering, Triangles) and overarching centrality metrics (BetweennessCentrality) suggest that nodes embedded within densely interconnected groups can assume significant roles on a broader network scale. The consistent negative correlations between "Capital" and other network metrics might indicate that resources (as defined by "Capital") are not necessarily held or controlled by the most connected or central nodes. This observation highlights a key characteristic of the automotive SoC supply chain network: companies with substantial capital may provide services or products to a select group of enterprises consistently, while the majority of nodes with numerous connections might possess only moderate resources. This phenomenon suggests a strategic distribution of resources or a balancing mechanism within the network, where nodes positioned less centrally hold significant resources, which helps mitigating risks or decentralizing control.

**4.4 Interaction and Disruption Analysis**

The typological structure of supply chain greatly determines the impact degree of disruption, therefore, supply chain structure parameters are considered to have a crucial impact on the resilience of the entire supply chain[55–57]. Through stimulating interaction and disruption process in Chinese automotive SoC supply chain network, we find that the percent of disruption in the supply chain will increase slower if enterprises have a strong self-recover ability, which enlighten the firms to strength their initiative ability. Figure 9 illustrates the proportion of affected firm nodes under HAD, HCA and HIA strategy in RC-IDM, the initial parameters are set as $\lambda = 0.5$, $\mu = 0.1$.

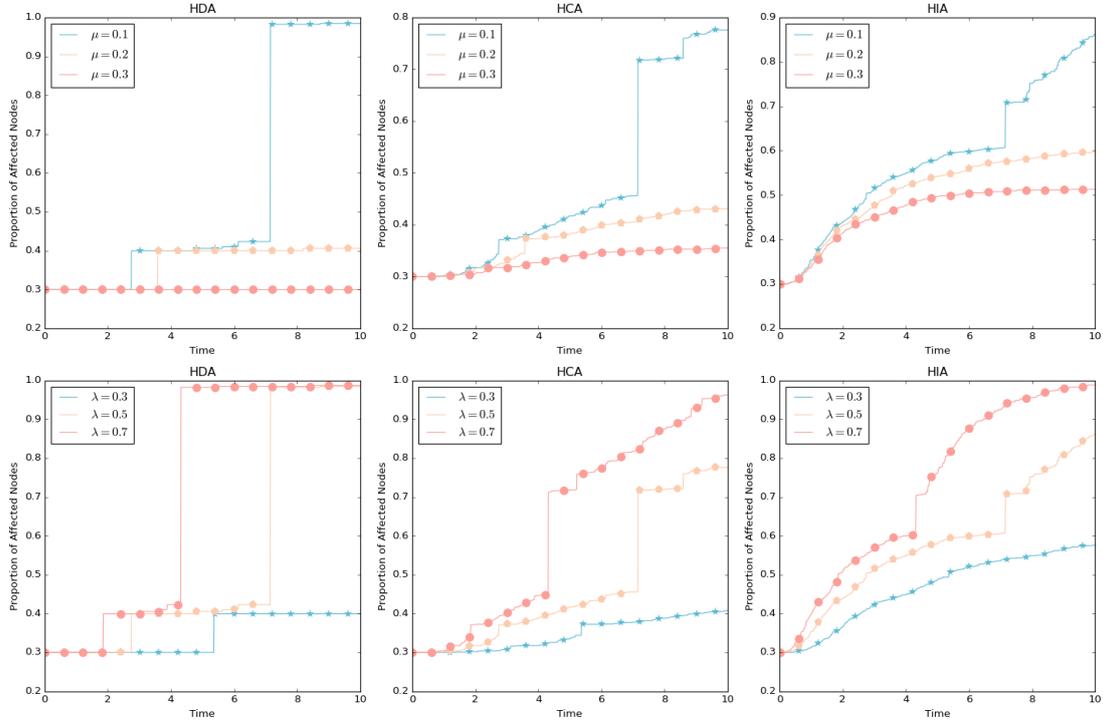

Fig.10 The proportion of affected firm nodes under HAD, HCA and HIA in RC-IDM.

If the enterprises have strong innovation capabilities, then in the event of a disruption, the whole supply chain will not only avoid being compromised, but it may even recover and thrive with a stable state. As shown in Figure 10, We calculate the critical recovery ratio by dividing $\mu$ by $\lambda$, find the whole automotive SoC supply chain could maintain function well if critical recovery ratio is more than 0.4. This gives a significant insight that anti-risk capabilities enable the entire supply chain to not only avoid detrimental impacts but also to potentially recover and thrive in the event of an unexpected setback, such as supply shortages or technological changes. Enhanced anti-risk ability leads to greater adaptability and resilience, allowing these enterprises to quickly devise and implement effective solutions. Consequently, the supply chain can maintain or even improve its efficiency and stability. This resilience can manifest in faster recovery times, the introduction of new and improved products, and stronger competitive positioning in the market. Thus, a focus on strengthening innovation capacities is crucial for sustaining long-term health and success in the dynamic environment of the automotive SoC industry.

We also observed that a sudden transition often occurs within the cascade interruption process of the entire supply chain network under the HDA and HCA strategies, particularly when there is a significant disparity between the destructive capabilities of larger neighbor firm nodes and the recovery abilities of smaller firm nodes. This indicates the presence of a critical threshold in the automotive SoC supply chain. Before this threshold, the overall damage to the network remains relatively minor. However, the network rapidly approaches collapse once passing this point. These insights are crucial for policymakers to regulate the entire automotive SoC supply chain at a macro level to prevent network collapse. Additionally, at a micro level, enterprises can enhance their resilience by bolstering their innovation capacities and engaging in horizontal or vertical alliances, thereby reducing dependence on their upstream and downstream supply chain partners and better mitigating potential disruptions throughout the supply chain. In Figure 7, there a weak correlation between corporate capital and certain characteristics of enterprise nodes. Under the HIA strategy, the propagation of risk across enterprises is relatively stable, with less pronounced sudden changes at the critical point. This aligns with the scale-free nature of the automotive SoC supply chain, where a few enterprises with substantial capital occupy significant positions, while the importance of other enterprises tends to be similar.

In RT-IDM, we compare the ratio of disrupted nodes to total nodes across different attack strategies. We find that when enterprises adopt absorb strategy — actively managing risk from supply chain partners rather than dispersing it among the remaining partners — the disruption process within the entire supply chain slows down. Conversely, if the enterprise transfers risk to its partners, the supply chain quickly reaches a state of paralysis. Under a random strategy, where each surviving enterprise has a 50% chance of choosing absorb strategy and the remaining probability of choosing transfer strategy, the speed of disruption in the supply chain network is relatively mitigated.

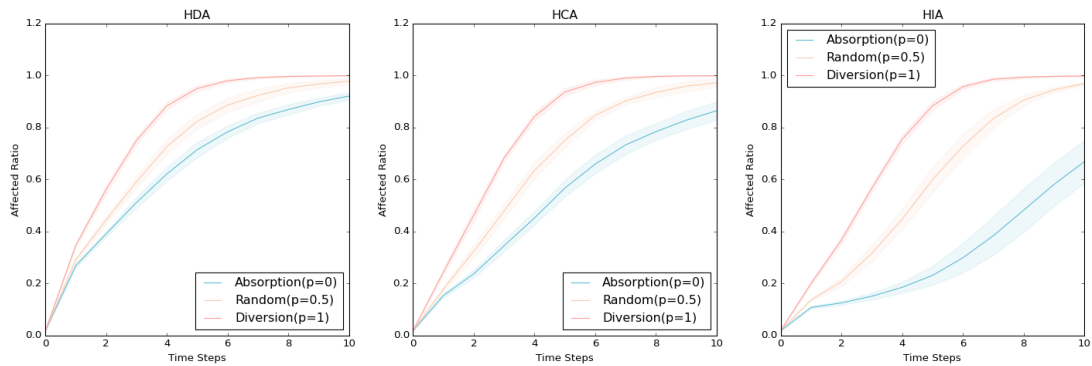

Fig.11 The proportion of affected firm nodes under community leader HAD, HCA and HIA in RT-IDM.

When we focus on the changes of registered capital of all firm nodes within supply chain network, some interesting phenomena emerge. Under three different attack strategies, the global trend in registered capital changes is similar. The global registered capital of the entire network decreases rapidly in the short term once absorb strategy is chosen, but as time progresses, the rate of decrease gradually slows down. On the other hand, although the initial rate of decrease is slower under transfer strategy, the rate of decline accelerates significantly in the mid-term. This indicates that with transfer strategy, as cascading risks propagate, the collapse speed of the registered capital of entire supply chain network also increases rapidly. As shown in Figure 12.

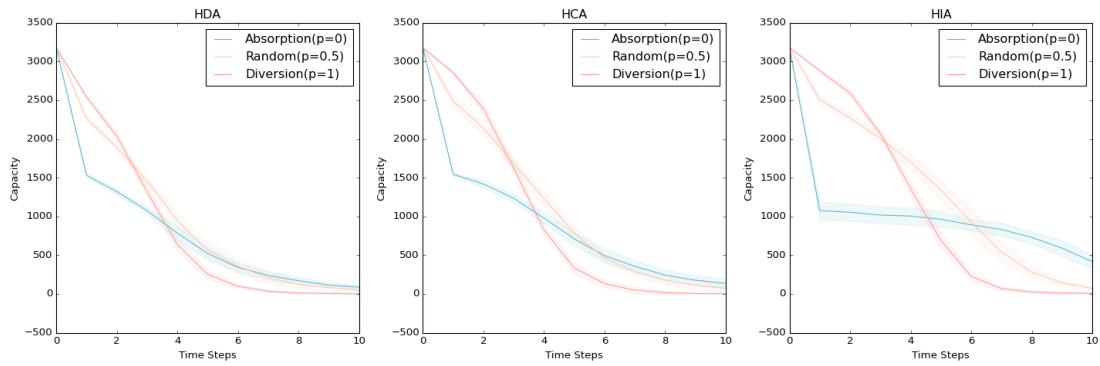

Fig.12 The proportion of automotive SoC supply chain capacity under community leader HAD, HCA and HIA strategy in CPDM.

The analysis above enlightens us that enterprises in Chinese automotive SoC industry field should consider absorbing a portion of the risks themselves rather than simply transferring them to other supply chain partners under specific situation. This approach can enhance the inherent resilience and stability of the entire supply chain, thereby reducing the risk of disruptions when external shocks happened.

## 5. Conclusion and Discussion

Sharing supply chain information across organizational boundaries in a rational and effective manner is crucial for enhancing enterprise decision-making and collaboration ability, especially in some specific industry fields such as automotive SoC. There are few studies to explore the collection and process of the Chinese supply chain data on open domain data, and most supplier data in professional database exists data lacking and anonymous supplier problem. Consequently, numerous studies rely on simplified and assumption-based simulated supply chain models, which struggle to completely encompass the complex dynamics, unpredictability, and variability of actual supply chains. This often leads to a disparity between theoretical results derived from simulations and the practical outcomes seen in real-world phenomenon. Open domain and professional database contain massive corpus related to supply chain knowledge including the firm name and the cooperative relationship between two enterprises, which are appropriate data source for named entity recognition and relation complement task. Overall, many studies overlook the relationship between real-world data and synthetic data. Even when utilizing real-world data, most is extracted directly from professional financial or stock databases. However, issues such as data scarcity, anonymous suppliers persist, and there remains a lack of truly effective methods for collecting real-world supply chain data and simulating interactions and disruptions.

In this research, we introduce an OSKEC-IDM framework for supply chain network construction and cascade failure simulation amidst limited information availability. Empirical Chinese automotive SoC supply chain triplets are obtained from both open domain and professional database. In detail, we utilize NER and data scraping techniques to gather crucial data from industry-leading websites, CSMAR, and Wind. This data is used to create a foundational dataset of firm entities. Then cross-domain transfer learning is used to mitigate data scarcity, KG and Levenshtein distance calculations for relationship mapping and entity similarity assessments. The connections between these entities are enhanced using inference technology, which aids in expanding the network scale of the supply chain. The characteristics of the supply chain network are then analyzed. Following this, two Interaction Models, namely the RC-IDM) and RT-IDM, are developed to simulate the complex behaviors of Chinese automotive SoC enterprises.

This research contributes to the understanding the structure of supply chain networks, particularly in contexts of scarce resources. Using the Chinese automotive SoC industry as a case study, we observe that its supply chain network demonstrates scale-free characteristics. Additionally, there is a noticeable similarity between the distribution of network degrees and registered capital. We further analyze the correlation among classic network indices and delve into the reasons why certain indices display either weak or strong correlations. This analysis helps to elucidate the underlying structural and functional dynamics of the Chinese automotive SoC supply chain network. In RC-IDM, to counteract the effects of disruption spread, an enterprise can bolster its anti-risk capabilities to maintain global stability within the real-world network. Furthermore, it is identified that the critical ratio of recovery ability to the influence of neighboring firm nodes is 0.4. This ratio is essential for ensuring that the government could effectively manage disruptions and sustain the industry supply chain operations. Subsequently, we discuss the proactive strategies that enterprises could adopt to bolster resilience of the whole supply chain against disruptions. When the risk absorption strategy is adopted, by proactively managing risks from supply chain partners instead of distributing them among other partners, the disruption process within the entire supply chain decelerates, and the global registered capital experiences a rapid decline in the short term, but this rate of decrease gradually moderates over time. In contrast, if an enterprise opts to transfer risk to its partners, the supply chain rapidly approaches a state of paralysis; meanwhile, the initial decrease in registered capital occurs at a slower pace, but the rate of decline significantly accelerates during the mid-term.

The findings of this study underscore the critical importance of risk management strategies, especially in complex and interdependent supply chain environments. Such results offer a fresh perspective on supply chain management, particularly in formulating more effective risk management and mitigation strategies. By implementing more flexible and responsive risk management strategies within the supply chain, enterprises can not only protect themselves from severe financial losses but also contribute to the stability and development of the entire industry. Future research should further explore how to optimize risk-absorption and risk-sharing strategies to enhance the supply chain's adaptability to future uncertainties.

However, there are significant challenges in obtaining precise data due to the dynamic nature of supply chain information. We must acknowledge that it is not feasible to capture all supply cooperative data from open domains or professional databases but we could the data as much as possible to describe the real-world collaboration. Additionally, the IDM model needs to incorporate more real-world scenarios and provide a comprehensive analysis of the automotive SoC supply chain network to enhance its effectiveness and relevance.

All in all, this study not only opens new vistas for Automotive SoC supply chain management, providing a universal strategy for constructing supply chain network but also lays a scientific foundation for assessing the resilience of supply chain networks under cascade failure impacts.

**Acknowledgements**

We are grateful to Wanqiu Cheng, Xiaopeng Li, and Xiangyun Meng for their assistance in developing OSKEC approach. We also thank Bitao Dai for his invaluable insights to the construction of the interaction disruption model. This work was supported by the National Natural Science Foundation of China (72025405, 72088101, 72001211, 72301285), the National Social Science Foundation of China (22ZDA102), the Hunan Science and Technology Plan Project (2020TP1013), the Natural Science Foundation of Hunan Province (2023JJ40685), the Innovation Team Project of Colleges in Guangdong Province (2020KCXTD040), the National Postdoctoral Program for Innovative Talents of China (BX20230475), and the Postgraduate Scientific Research Innovation Project of Hunan Province (XJCX2023163). The authors declare that they have no conflict of interest. (*Corresponding author: Xin Lu*)

# Reference

[1] Ascari G, Bonam D, Smadu A. Global supply chain pressures, inflation, and implications for monetary policy[J]. Journal of International Money and Finance, 2024, 142: 103029.

[2] Kang Zhao, Zhiya Zuo, Jennifer Blackhurst. Modelling supply chain adaptation for disruptions: An empirically grounded complex adaptive systems approach[J]. Journal of Operations Management, 2019, 65: 190-212.

[3] Niklas Berger, Stefan Schulze-Schwering, Elisa Long, et al. Risk management of supply chain disruptions: An epidemic modeling approach[J]. European Journal of Operational Research, 2023, 304(3): 1036-1051.

[4] Cohen P. Malaysia Rises as Crucial Link in Chip Supply Chain[N]. The New York Times, 2024-03-13.

[5] Mickle T, McCabe D, Swanson A. How the Big Chip Makers Are Pushing Back on Biden's China Agenda[N]. The New York Times, 2023-10-05.

[6] Marra M, Ho W, Edwards J S. Supply chain knowledge management: A literature review[J]. Expert Systems with Applications, 2012, 39(5): 6103-6110.

[7] Cerchione R, Esposito E. A systematic review of supply chain knowledge management research: State of the art and research opportunities[J]. International Journal of Production Economics, 2016, 182: 276-292.

[8] Pathak S D, Day J M, Nair A, et al. Complexity and Adaptivity in Supply Networks: Building Supply Network Theory Using a Complex Adaptive Systems Perspective*[J]. Decision Sciences, 2007, 38(4): 547-580.

[9] Anderson P. Perspective: Complexity Theory and Organization Science[J]. Organization Science, 1999, 10(3): 216-232.

[10] Choi T Y, Dooley K J, Rungtusanatham M. Supply networks and complex adaptive systems: control versus emergence[J]. Journal of Operations Management, 2001, 19(3): 351-366.

[11] Sarimveis H, Patrinos P, Tarantilis C D, et al. Dynamic modeling and control of supply chain systems: A review[J]. Computers & Operations Research, 2008, 35(11): 3530-3561.

[12] Piya S, Shamsuzzoha A, Khadem M. An approach for analysing supply chain complexity drivers through interpretive structural modelling[J]. International Journal of Logistics Research and Applications, 2020, 23(4): 311-336.

[13] Xu N R, Liu J B, Li D X, et al. Research on Evolutionary Mechanism of Agile Supply Chain Network via Complex Network Theory[J]. Mathematical Problems in Engineering, 2016, 2016: e4346580.

[14] HP T, Raghavan U N, Kumara S, et al. Survivability of multiagent-based supply networks: a topological perspect[J]. IEEE Intelligent Systems, 2004, 19(5): 24-31.

[15] Helbing D, Armbruster D, Mikhailov A S, et al. Information and material flows in complex networks[J]. Physica A: Statistical Mechanics and its Applications, 2006, 363(1): xi-xvi.

[16] Pathak S D, Dilts D M, Biswas G. On the evolutionary dynamics of supply network topologies[J]. IEEE Transactions on Engineering Management, 2007, 54(4): 662-672.

[17] Li Y, Yuan Y. Supply chain disruption recovery strategies for measuring profitability and resilience in supply and demand disruption scenarios[J]. RAIRO - Operations Research, 2024, 58(1): 591-612.

[18] Pathak S D, Dilts D M, Mahadevan S. Investigating Population and Topological Evolution in a Complex Adaptive Supply Network*[J]. Journal of Supply Chain Management, 2009, 45(3):

54-57.

[19] Zeng Y, Xiao R. Modelling of cluster supply network with cascading failure spread and its vulnerability analysis[J]. International Journal of Production Research, 2014, 52(23): 6938-6953.

[20] Kito T, New S, Ueda K. How automobile parts supply network structures may reflect the diversity of product characteristics and suppliers' production strategies[J]. CIRP Annals, 2015, 64(1): 423-426.

[21] Kito T, Ueda K. The implications of automobile parts supply network structures: A complex network approach[J]. CIRP Annals, 2014, 63(1): 393-396.

[22] Brintrup A, Wang Y, Tiwari A. Supply Networks as Complex Systems: A Network-Science-Based Characterization[J]. IEEE Systems Journal, 2017, 11(4): 2170-2181.

[23] Xin Lu, Abigail L. Horn, Jiahao Su, et al. A Universal Measure for Network Traceability[J]. Omega, 2019, 87: 191-204.

[24] Cai M, Huang G, Tan Y, et al. Decoding the complexity of large-scale pork supply chain networks in China[J]. Industrial Management & Data Systems, 2020, 120(8): 1483-1500.

[25] Calatayud A, Mangan J, Palacin R. Vulnerability of international freight flows to shipping network disruptions: A multiplex network perspective[J]. Transportation Research Part E: Logistics and Transportation Review, 2017, 108: 195-208.

[26] Gualandris J, Longoni A, Luzzini D, et al. The association between supply chain structure and transparency: A large-scale empirical study[J]. Journal of Operations Management, 2021, 67(7): 803-827.

[27] Spiegler V L M, Naim M M, Towill D R, et al. A technique to develop simplified and linearised models of complex dynamic supply chain systems[J]. European Journal of Operational Research, 2016, 251(3): 888-903.

[28] Lou Y, Wang L, Chen G. Structural Robustness of Complex Networks: A Survey of A Posteriori Measures[Feature][J]. IEEE Circuits and Systems Magazine, 2023, 23(1): 12-35.

[29] Kim Y, Chen Y S, Linderman K. Supply network disruption and resilience: A network structural perspective[J]. Journal of Operations Management, 2015, 33-34: 43-59.

[30] Nair A, Narasimhan R, Choi T Y. Supply Networks as a Complex Adaptive System: Toward Simulation-Based Theory Building on Evolutionary Decision Making[J]. Decision Sciences, 2009, 40(4): 783-815.

[31] Son B G, Chae S, Kocabasoglu-Hillmer C. Catastrophic supply chain disruptions and supply network changes: a study of the 2011 Japanese earthquake[J]. International Journal of Operations & Production Management, 2021, 41(6): 781-804.

[32] Cheng A L, Fuchs E R H, Karplus V J, et al. Electric vehicle battery chemistry affects supply chain disruption vulnerabilities[J]. Nature Communications, 2024, 15(1): 2143.

[33] Wang Q, Zhou H, Zhao X. The role of supply chain diversification in mitigating the negative effects of supply chain disruptions in COVID-19[J]. International Journal of Operations & Production Management, 2023, 44(1): 99-132.

[34] Takawira B, Pooe R I D. Supply chain disruptions during COVID-19 pandemic: Key lessons from the pharmaceutical industry[J]. South African Journal of Business Management, 55(1): 4048.

[35] Ficara A, Curreri F, Fiumara G, et al. Human and Social Capital Strategies for Mafia Network Disruption[J]. IEEE Transactions on Information Forensics and Security, 2023, 18: 1926-1936.

[36] Tagliari M M, Bogoni J A, Blanco G D, et al. Disrupting a socio-ecological system: could traditional ecological knowledge be the key to preserving the Araucaria Forest in Brazil under climate change?[J]. Climatic Change, 2023, 176(2): 2.

[37] Ki D H, Yu H, Kim D, et al. Discovery of a potent, selective, and orally available small molecule for disruption of the SOS1-RAS interaction[J]. Cancer Research, 2024, 84(6_Supplement): 3313-3313.

[38] Wang S, Sun X, Li X, et al. GPT-NER: Named Entity Recognition via Large Language Models[A]. arXiv, 2023.

[39] Devlin J, Chang M W, Lee K, et al. BERT: Pre-training of Deep Bidirectional Transformers for Language Understanding[A]. arXiv, 2019.

[40] Torres D V, Freitag M, Cherry C, et al. Prompting PaLM for Translation: Assessing Strategies and Performance[J]. 2023.

[41] Sabane M, Ranade A, Litake O, et al. Enhancing Low Resource NER using Assisting Language and Transfer Learning[C]//2023 2nd International Conference on Applied Artificial Intelligence and Computing (ICAAIC). 2023: 1666-1671.

[42] Hu Y, Chen Q, Du J, et al. Improving large language models for clinical named entity recognition via prompt engineering[J]. Journal of the American Medical Informatics Association, 2024: ocad259.

[43] Oliveira V, Nogueira G, Faleiros T, et al. Combining prompt-based language models and weak supervision for labeling named entity recognition on legal documents[J]. Artificial Intelligence and Law, 2024.

[44] Wang L, Zhao W, Wei Z, et al. SimKGC: Simple Contrastive Knowledge Graph Completion with Pre-trained Language Models[A]. arXiv, 2022.

[45] Sun J, Xu C, Tang L, et al. Think-on-Graph: Deep and Responsible Reasoning of Large Language Model on Knowledge Graph[A]. arXiv, 2023.

[46] Saisridhar P, Thürer M, Avittathur B. Assessing supply chain responsiveness, resilience and robustness (Triple-R) by computer simulation: a systematic review of the literature[J]. International Journal of Production Research, 2024, 62(4): 1458-1488.

[47] Bai X, Ma Z, Zhou Y. Data-driven static and dynamic resilience assessment of the global liner shipping network[J]. Transportation Research Part E: Logistics and Transportation Review, 2023, 170: 103016.

[48] Guntuka L, Corsi T M, Cantor D E. Recovery from plant-level supply chain disruptions: supply chain complexity and business continuity management[J]. International Journal of Operations & Production Management, 2023, 44(1): 1-31.

[49] Blondel V D, Guillaume J L, Lambiotte R, et al. Fast unfolding of communities in large networks[J]. Journal of Statistical Mechanics: Theory and Experiment, 2008, 2008(10): P10008.

[50] Elsler L G, Oostdijk M, Gephart J A, et al. Global trade network patterns are coupled to fisheries sustainability[J]. PNAS Nexus, 2023, 2(10): pgad301.

[51] Alexander M, Forastiere L, Gupta S, et al. Algorithms for seeding social networks can enhance the adoption of a public health intervention in urban India[J]. Proceedings of the National Academy of Sciences, 2022, 119(30): e2120742119.

[52] Goh K I, Cusick M E, Valle D, et al. The human disease network[J]. Proceedings of the National Academy of Sciences of the United States of America, 2007, 104(21): 8685-8690.

[53] Barabási A L. The origin of bursts and heavy tails in human dynamics[J]. Nature, 2005,

435(7039): 207-211.

[54] Kim S J, Lee S H. An Improved Computation of the PageRank Algorithm[M]//Crestani F, Girolami M, Van Rijsbergen C J. Advances in Information Retrieval: Vol. 2291. Berlin, Heidelberg: Springer Berlin Heidelberg, 2002: 73-85.

[55] Ivanov D. Revealing interfaces of supply chain resilience and sustainability: a simulation study[J]. International Journal of Production Research, 2018, 56(10): 3507-3523.

[56] Ivanov D, Dolgui A, Sokolov B, et al. Literature review on disruption recovery in the supply chain*[J]. International Journal of Production Research, 2017, 55(20): 6158-6174.

[57] Pavlov A, Ivanov D, Dolgui A, et al. Hybrid fuzzy-probabilistic approach to supply chain resilience assessment[J]. IEEE Transactions on Engineering Management, 2018, 65(2): 303-315.